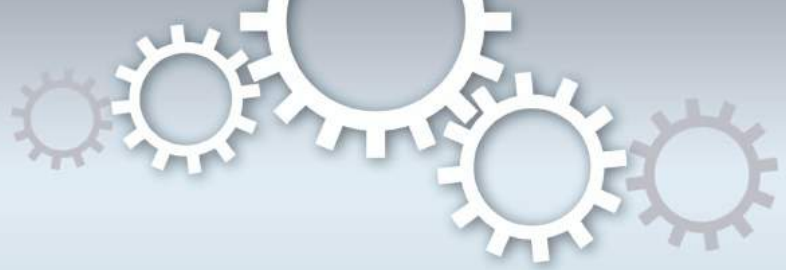

# Introduction of impermeable actin-staining molecules to mammalian cells by optoporation

Kamal Dhakal, Bryan Black & Samarendra Mohanty

Biophysics and Physiology Lab, Department of Physics, University of Texas at Arlington, Texas, USA.

The selective insertion of foreign materials, such as fluorescent markers or plasmids, into living cells has been a challenging problem in cell biology due to the cell membrane's selective permeability. However, it is often necessary that researchers insert such materials into cells for various dynamical and/or drug delivery studies. This problem becomes even more challenging if the study is to be limited to specific cells within a larger population, since other transfection methods, such as viral transfection and lipofection, are not realizable with a high degree of spatial selectivity. Here, we have used a focused femtosecond laser beam to create a small transient hole in the cellular membrane (optoporation) in order to inject nanomolar concentrations of rhodamine phalloidin (an impermeable dye molecule for staining filamentous actin) into targeted living mammalian cells (both HEK and primary cortical neurons). Following optoporation, the dye bound to the intracellular actin network and rise in fluorescence intensity was observed. Theoretical dynamics of the dye's diffusion is discussed, and numerical simulations of diffusion time constants are found to match well with experimental values.

I t is necessary in many biological and medical studies to introduce fluorescent markers or plasmids into cells to facilitate visualization of filamentous actin, tubulin, or to observe functional alteration[1] of cellular processes. Though cells can be fixed and fluorescently stained with cytoskeletal dyes, there has always been significant interest in the visualization of cytoskeletal dynamics within living cells. Several methods have been developed to accomplish this aim. Among them, one widely used method is viral transfection of actin-GFP plasmids. However, viral transfection methods cause expression in large populations of cells. Several physical methods have also been developed for localized injection of exogenous factors and visualization of cytoskeletal dynamics, with the two most widely employed methods being electroporation[2] and microinjection[3]. These methods are either highly invasive, increase the possibility of contamination, or lack spatial and temporal specificity. For example, micro-injection frequently causes unintended cell damage, whereas electroporation has been reported to achieve relatively low efficiencies[4]. Chemical methods, such as lipofection and DEAE-dextran, have also been employed for the introduction of exogenous factors[5–6]. These methods may also lead to adverse effects[7], do not allow for single cell specificity, and are currently only relevant in large populations of cells. With the advent of laser technology, light has been increasingly utilized for safe and sterile biological applications, such as micro-surgery[8] and single cell manipulation[9–11].

Optoporation (or photoporation) is an optical method of creating a transient hole in the cell membrane or organelle by briefly (on the scale of milliseconds) exposing it to a focused pulsed laser of sufficient intensity[12]. In the case of *in vitro* studies, this method allows for single cells to be targeted, and may be interfaced with microfluidic devices[10] or additional, fast-scanning optics for relatively high-throughput studies[13–14]. Furthermore, since optoporation is a non-contact method, and since it can take place within a completely contained system, it is absolutely sterile.

The primary mechanism by which a cellular membrane is optoporated varies depending on pulse width and laser intensity. Many hypotheses have been proposed, such as plasma formation, dielectric breakdown, cavity formation, local heating, shock wave formation[15–16], and photo-acoustic effects[17–18]. Due to use of a femtosecond pulsing laser beam, we have assumed that dielectric breakdown is the primary cause of the observed optoporation effect. The size of the hole which is formed during laser irradiance is dependent upon the pulse-energy, pulse-width, and wavelength of the light being used. Since the diffraction-limited spot size is directly proportional to the wavelength of light, ultra-violet (UV) wavelengths may seem to be better-suited over other wavelengths in near-infrared (NIR) regimes for optoporation. However, UV light carries a high risk of damage to the cells since it is highly absorbed by the cellular components. Therefore, the use of UV light may not be suitable[19–20]. CW argon lasers operating at 488 nm, or Nd: YAG lasers which operate at 532 nm, have also been employed for optoporation. However, many cell organelles have high absorption at these wavelengths, so the possibility of detrimental affects at these, and others wavelengths in the visible range, cannot be ruled out. It is well known that the absorption of light by intracellular components is low in the NIR (700–1100 nm) range. This range of wavelengths is, in fact, often referred to as the 'therapeutic window' for their ability to probe biological samples with low deleterious effects. We have chosen to use wavelengths in the range of 800–850 nm, where our laser has the lowest pulse width (~100 fs) and highest repetition rate (~80 MHz)[21].

Here, we report use of ultrafast laser pulses to create a single transient hole in the cellular membrane in order to inject nanomolar concentrations of rhodamine phalloidin (RP, an impermeable filamentous actin dye molecule) into single viable mammalian cells (both HEK and primary cortical neurons). Theoretical dynamics of the dye's diffusion is discussed, and numerical simulations of diffusion time constants are found to match well with experimental values.

## Results and Discussion

**Optimization of dye concentration and laser parameters.** All experiments and optimization of dye and laser parameters were carried out on the same experimental setup (Fig. 1 (a)). In order to optoporate efficiently, the focused laser beam should be symmetric and share a focal plane with the sample's surface. This can be verified by beam symmetry as well as circular rings around the spot which is shown in Fig. 1 (b–c). Optimization of the laser parameters was undertaken by varying the laser power and exposure time (macropulse duration, 20–50 ms), while keeping the wavelength constant (800 nm). At higher average laser powers (at or above 130 mW) and macro-pulse durations (above 30 ms), HEK cells were observed to die (quick rise in RP-fluorescence and significant change in morphology) as seen in Suppl. Fig. 1. It is known that the Phalloidin family of dyes are toxic to living cells, and can cause or contribute to cell death[22] at high concentrations. Therefore, imaging is typically performed on fixed cells to visualize the cytoskeletal filamentous actin network. Here, we have determined a window of optimized RP dye concentration by monitoring the cell viability (by morphology and calcein exclusion assay) and RP-fluorescence following optoporation. In Suppl. Fig. 2, we show Calcein-AM images of the cells following optoporation. Optical parameters were kept constant while the RP concentration was varied from 40 to 200 nM. At lower concentrations (42 nM), we did not observe any change in fluorescence signal. At higher RP concentration (168 nM), optoporated cells showed a sharp increase in fluorescence immediately following optoporation which was not attributed to RP dye binding, along with notable changes in morphology (Fig. 2 (e–h)), indicating cell death. Since the cell viability was retained following optoporation and injection of 126 nM concentration of RP, live cell imaging of the cytoskeletal actin network could be performed (Fig. 1 (d)).

**Femtosecond laser-assisted targeted dye injection.** Optimized parameters (laser power, exposure time, and RP concentration) were used for optoporation into HEK and rat cortical neurons. First, RP was optically injected into HEK cells. There was a clear, observable increase in the cell's fluorescence (Fig. 2 (a–d)) over a time period of approximately 10 minutes. The fluorescence signal increased slowly and ultimately approached a saturation level (independent of quenching) as predicted by Fick's law of diffusion (discussed in the following sections). As can be clearly seen in Fig. 2 (c), the most intense fluorescence is observed along the periphery of the cell's membrane, where the filamentous actin network is most dense. Next, we demonstrate the introduction of RP by optoporation into viable primary embryonic rat cortical neurons (RCNs). Fig. 2 (i–p) shows the significant increase in fluorescence in the cell body as well as neurites of the RCN. The live polymerization of actin in the growth cone of the neuron is shown by the dynamics of the RP-stained fluorescent actin (Suppl. Movie 1). Hence, with optimized laser parameters and dye concentration, cell-impermeable RP-dye could be successfully injected to the individually targeted cells.

The selective targeting of cells (e.g. cancers, neurons) of individual cell type or spatially-targeted cells is critical to studies of drug delivery, function, and therapeutics. For example, the brain contains a large network of neurons, astrocytes, etc., the organization and dynamics of which characterize and control bodily function. These cells form physico- chemical connections with one another and form an elaborate circuitry. In order to study the connections and com-munication within these cells, it may often prove necessary to inject dyes, opsins[11,23], or plasmids[24]. This can be achieved by femtosecond laser–based cellular optoporation. By optoporating a single cell (or small subpopulation of cells) among a larger network of neurons, researchers will be able to distinguish individual components and functions in neural networks. Fig. 3 (a) shows the non-uniform distribution of fluorescence intensity across line drawn in Fig. 3 (b). The fluorescence intensity is at maximum near the optoporation site (center of cell body) and decreases along the periphery of the cell body (Suppl. Movie 2). Fig. 3 (c, d) show the bright field and fluorescence images before optoporation. Fig. 3 (e, f), show the bright field and fluorescence images after the optoporation. Fig. 3 (f) shows an optoporated cell whose components were fluorescent in an intermingled network, illustrating the potential of optoporation for studying single-cell components of larger cellular networks.

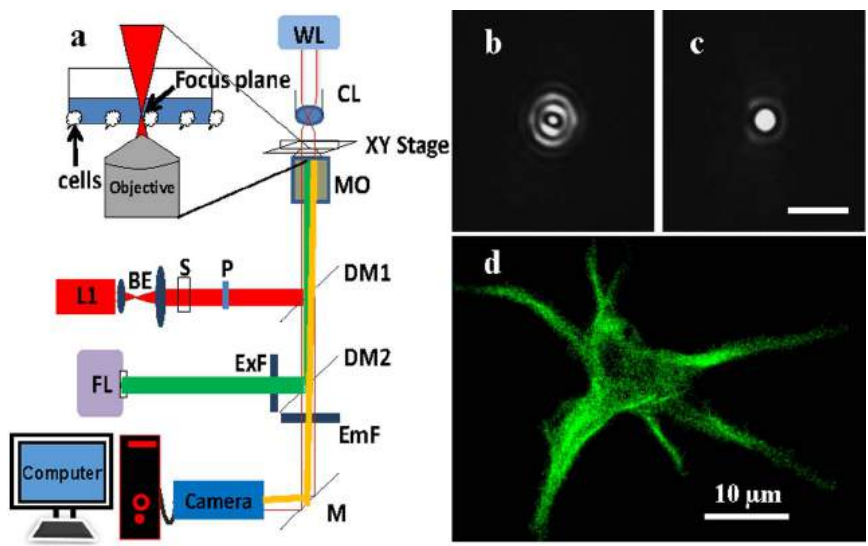

**Figure 1** | (a) Experimental setup for femtosecond laser-assisted optoporation. L: Ti: Sapphire laser; BE: beam expander; S: shutter; P: polarizer; Fl.: fluorescence excitation source; Ex: excitation filter; Em: emission filter; MO: microscope objective; CL: condenser lens; DM1 and DM2: dichroic mirrors; M: mirror; WL: halogen lamp. (b–c) Symmetric laser spot focused above and at the sample's surface plane. (d) Rat cortical neuron approximately 30 minutes following fs laser optoporation and injection with rhodamine phalloidin. (This figure was drawn by author (KD) in Microsoft PowerPoint software.).

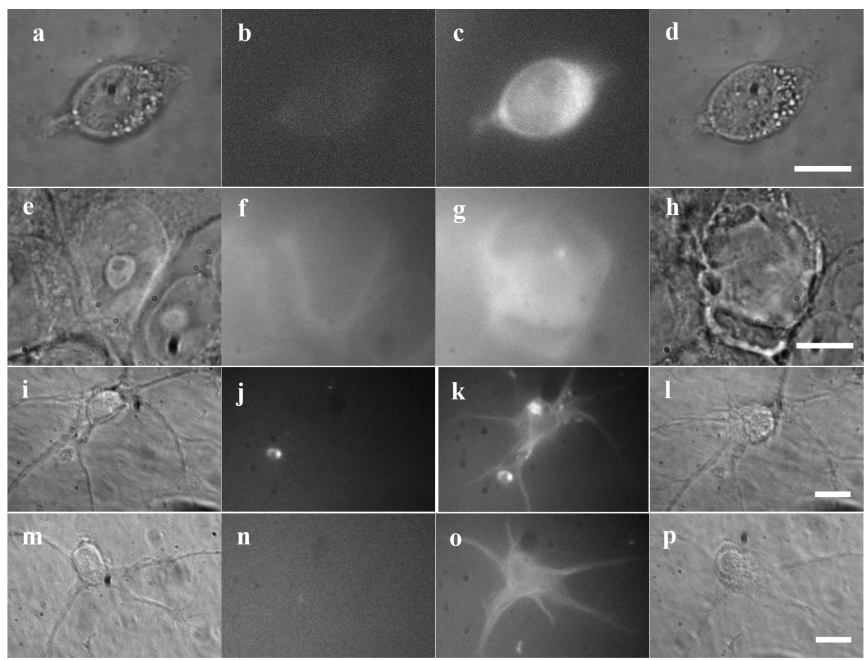

**Figure 2 | Images of HEK and rat cortical neurons before (bright field, then fluorescence) and after (fluorescence, then bright field) optoporation and injection with rhodamine phalloidin.** (a–d) HEK, 126 nM RP concentration. (e–h) HEK, 168 nM RP concentration. (i–l, m–p) Rat cortical neuron, 120 nM RP concentration.

Furthermore, following optical injection of RP, we were able to detect the motion of intracellular component or vesicle, which may have been a mitochondrial vesicle (Suppl. Fig. 3 and Suppl. Movie 3), since mitochondrial vesicles are known to be non-selectively stained by rhodamine phalloidin[25]. This illustrates the potential of this method for studying other intracellular mechanics besides actin network reorganization.

**Numerical simulation of laser ablation.** Many laser parameters, such as laser intensity, pulse width, repetition rate, and the focal volume (interaction volume) play important roles in laser-assisted optoporation[12]. Optoporation is only possible when the peak intensity reaches the threshold-value, producing a sufficient free electron density at the focal volume. We are considering that the mechanism of biological membrane ablation by femtosecond laser is primarily due to low density plasma formation at the focal volume[26]. The low density plasma is formed by multiphoton ionization process, often referred to as cascade or avalanche ionization. The energy of a single photon with wavelength of 800 nm is 1.56 eV. In order to cross the band gap, which can be considered as 6.5 eV (that of water) for cell culture medium, we need at least five photons to cause electron transitions to the excitation band. The critical free electron density (plasma) at the laser's focus, above which laser ablation starts, is defined by

$$Q_{cr} = \frac{\omega^2 \varepsilon_0 m_c}{e^2} \quad (1)$$

where $Q_{cr}$ is the critical plasma density, $\omega$ is plasma frequency, $\varepsilon_0$ is the dielectric constant in vacuum, and $m_c$ is the mass of the electron. At 800 nm wavelength, this corresponds to a $Q_{cr} = 1.8 \times 10^{21}$ cm$^{-3}$ and other investigators[27–28] have correlated this to a necessary power density of $1.3 - 2.6 \times 10^{13}$ W/cm$^2$. It is on this basis that we have chosen an experimental power density of $2.24 \times 10^{13}$ W/cm$^2$ during optoporation, which lies within the theoretically predicted limit.

The laser focal cross-section has been demonstrated to be an ellipsoid[29]. The short ($d$) and long dimensions ($l$) of the ellipsoid focal spot are given by the following relations[30], assuming a large solid angle.

$$d = \frac{1.22\lambda}{NA} \quad (2)$$

which results in a short ($d$) and long ($l$) axis value of 750 and 1800 nm, respectively.

$$\frac{l}{d} = \frac{1 - \cos\alpha}{\sqrt{3 - 2\cos\alpha - \cos\alpha}} \quad (3)$$

We have used the Gaussian distribution of laser irradiance at the focal volume with the above-mentioned major and minor axes to determine the ellipsoid distribution of irradiance.

$$I(r,z) = I(0,0) \exp\left(-2\left(\frac{r^2}{a^2} + \frac{z^2}{b^2}\right)\right) \quad (4)$$

where $r$ and $z$ are the coordinate in radial and axial distribution ($a = d/2$ and $b = l/2$).

The free electron density produced during the laser irradiation is proportional to the intensity $I^k$ [26], where k is the number of photons required to ionize the medium (k = 5, in our case).

$$Q(r,z) = \left[I(0,0) \exp\left(-2\left(\frac{r^2}{a^2} + \frac{z^2}{b^2}\right)\right)\right]^K \quad (5)$$

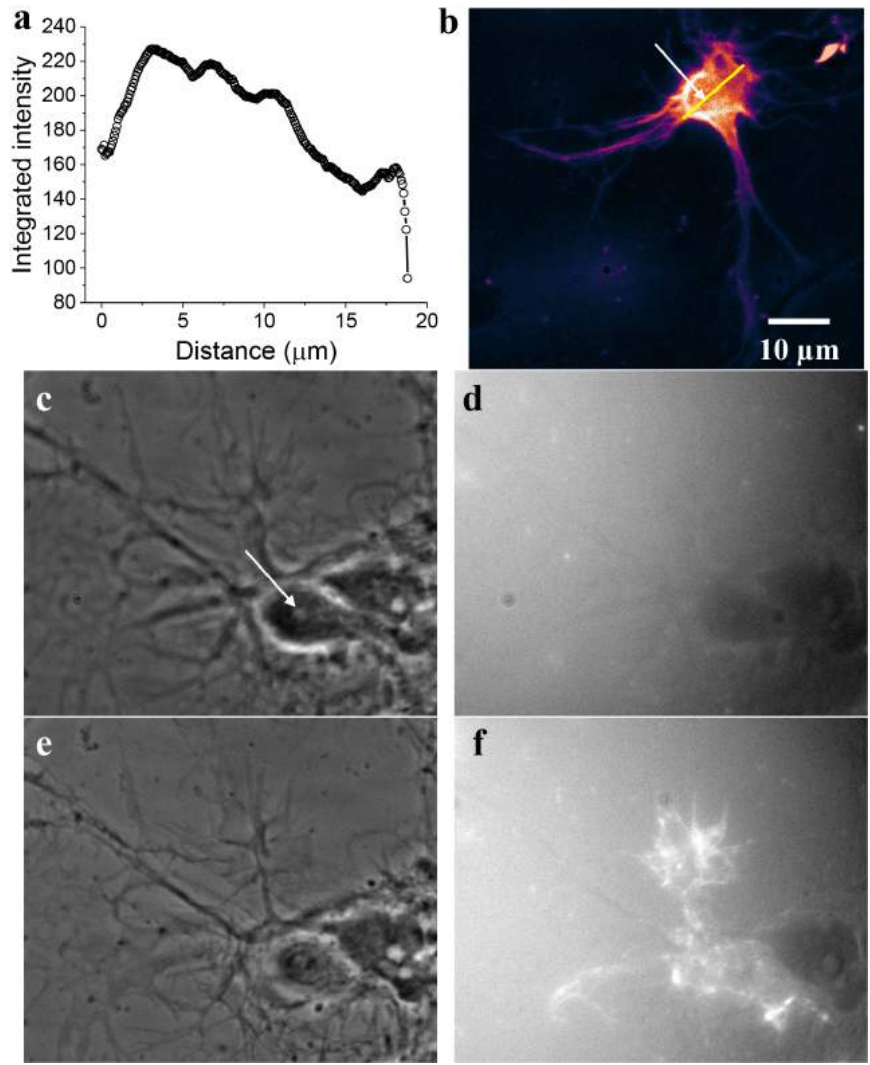

**Figure 3 | Non-uniform distribution of dye inside the rat cortical neuron cell body.** (a) Fluorescence intensity along the line (b) is maximum near the poration site and decreases away from the poration site. (c–d) Bright-field and fluorescence images before the optoporation, and (e–f) bright field and fluorescence images after the optoporation showing the targeted injection of dye. Arrow indicates the poration site.

$$Q(r,z) = Q(0,0) \exp\left(-2K\left(\frac{r^2}{a^2} + \frac{z^2}{b^2}\right)\right) \quad (6)$$

where $Q(0,0) = I(0,0)^k$ represents the free electron density at the center of the focal plane. Plots of equations 4 and 6 are shown in Figs. 4 (a and b). Experimentally, we have determined the pore's diameter to be approximately 1.6 μm, but theoretical calculations show that the ideal membrane pore might be significantly smaller than measured. Figs. 4 (b and c) show the pore as observed at the instance of pulsed laser irradiation during bright-field imaging. The parameters' values of physical quantities used in this study have been presented in Table 1.

**Dynamics of dye diffusion.** The dye diffusion time constant was experimentally determined by fitting the normalized fluorescence intensity in time (following optoporation) with Fick's law. Fick's first law of diffusion, $C - C_0 = D(dC/dt)$, a standard treatment for steady state diffusion, was used, where $C$ is the concentration of fluorescent dye at any instantaneous time, $C_0$ is the concentration of the dye in extracellular solution, and $D$ is the diffusion coefficient. Thus, the increase in fluorescence intensity could be fitted to a single exponential function as follows:

$C/C_0 = 1 - e^{-(t-t_0)/\tau}$, where $C/C_0$ is the normalized concentration of the fluorescent dye at time t, and $t_0$ is the initial time of dye influx with τ being the rise-time constant[21]. To calculate the diffusion coefficient and the diffusion time constant at different concentrations, we used the flux equation. The flux through a unit area can be calculated as

$$J(x,t) = C(x,t) \cdot v(x,t) \quad (7)$$

where C is the concentration of dye molecules contained in the injection volume and $v$ is the velocity of the dye.

The chemical potential is related with the concentration gradient by

$$\mu = \mu_0 + K_B T \cdot \ln\left(\frac{C}{C_0}\right) \quad (8)$$

where $K_B$ is Boltzmann constant. The velocity of the particle in the viscous medium is calculated by

$$\vec{v} = \sigma \vec{F} \quad (9)$$

where $\sigma = 1/6\pi\eta rv$, r being the molecule's radius of gyration and $F$ = viscous drag force which equals to potential gradient. Differentiating eq. (8)

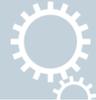

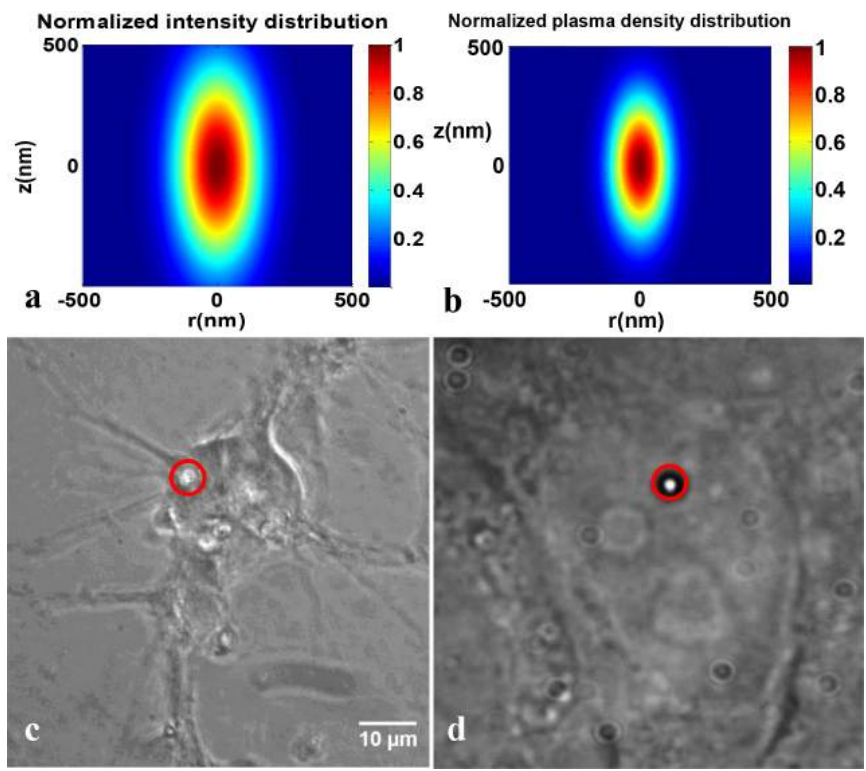

**Figure 4** | Theoretical intensity (a) and plasma distribution (b) of the fs-pulsed laser beam at the focal plane. A transient hole (marked by red circle) observed during optoporation (c–d).

$$F = -\nabla\mu = -k_B T \cdot \nabla C \quad (10)$$

From above equations,

$$J = -\frac{k_B T}{6\pi r \eta} \cdot \nabla C \quad (11)$$

Comparing with Fick's law, we see that

$$J = -D\nabla C \quad (12)$$

$$D = \frac{k_B T}{6\pi r \eta} \quad (13)$$

with $r$ = radius of gyration which is equivalent to $\sqrt{3/5}\,R$

The solution of equation (12) in one dimension can be written as

$$C = C_0 \left(1 - \exp\left(-\frac{t}{\tau}\right)\right) \quad (14)$$

$$\tau = \frac{t}{\ln\left(1 - \frac{C}{C_0}\right)} \quad (15)$$

Table 1 | Parameters used in dye diffusion model.

| | Definition | Units | Value |
|---|---|---|---|
| $\lambda$ | Laser wavelength | nm | 800 |
| $\varepsilon_0$ | Dielectric permittivity | F/m | $8.854 \times 10^{-12}$ |
| T | Temperature | K | 310 |
| $k_B$ | Boltzmann constant | J/K | $1.38 \times 10^{-23}$ |
| R | Radius of individual dye molecule | nm | 0.6997 |
| $\eta$ | Intracellular viscosity coefficient | cp | 2 |
| NA | Numerical aperture of microscope objective | - | 1.3 |
| $m_c$ | Mass of generated plasma | kg | $9.11 \times 10^{-31}$ |
| $\omega$ | Plasma frequency | $s^{-1}$ | $2.36 \times 10^{15}$ |

Figs. 5 (a) and (b) respectively show the increase in normalized fluorescence intensity following optoporation in the case of HEK cells and rat cortical neuron. The normalized fluorescence intensity as a function of RP concentration is shown in Fig. 5 (c). Fig. 5 (d) shows the comparison between these experimental values (blue line) and numerical simulations (red line) based on equation (14), where the time ($t$) was held fixed (40 sec). We have assumed diffusion of dye inside the cell following optoporation, so we have used the previously measured[31] intracellular viscosity coefficient, with the diffusion coefficient calculated by equation (13) and was found to be D = $2.13 \times 10^{-8}$ cm²/s, which matches with experimentally determined values of the diffusion coefficient of RP = $1.38 \times 10^{-8}$ cm²/s.

The targeted optical injection of rhodamine phalloidin into HEK cells and rat cortical neurons is an important step in the advancing technologies used to visualize intracellular actin networks. It is worth noting that rhodamine phalloidin remains one of the most trusted and widely used markers for f-actin networks *in vitro* due to its fluorescence efficiency and small relative size. Optical delivery of the dye or other external molecules in to the single cells by femtosecond laser beam offers significant advantages over other methods of injection/transfection such as micro-injection, electroporation, lipid based transfection, and viral transfection due to its low toxicity, spatial and temporal selectivity, and absolute sterility. However, there is great challenge to optoporate myriads of cells simultaneously without compromising the cell viability. Multiplexing the laser beam using spatial light modulator (SLM) and estimating proper laser parameters before conducting experiments will help to overcome this challenge.

## Methods

**Cell culture.** All experimental procedures were conducted according to UT Arlington Institutional Animal Care and Use Committee approved protocol. The cortical neurons were isolated from embryonic 18 day rat embryos. The cortical tissues were dissected, cleaned (meningeal layer), enzymatically dissociated (0.125% trypsin in L-15 medium) for 20 minutes at 37°C. The dissociated cortical neurons (100,000/device) were seeded on Poly-D-lysine (PDL, 0.01%, Sigma) pre-coated cover glass with Polydimethylsiloxane (PDMS) barrier well (Sylgard 184, Dow corning), and the serum-free culture medium (Neurobasal medium supplemented B-27 with BDNF and NT-3, 10 ng/ml) was changed every 3 days. The cell cultures were maintained at 37°C in a 5% $CO_2$, humidified atmosphere prior to experiments. The experiments are performed in an environment with controlled temperature, humidity, and $CO_2$ levels.

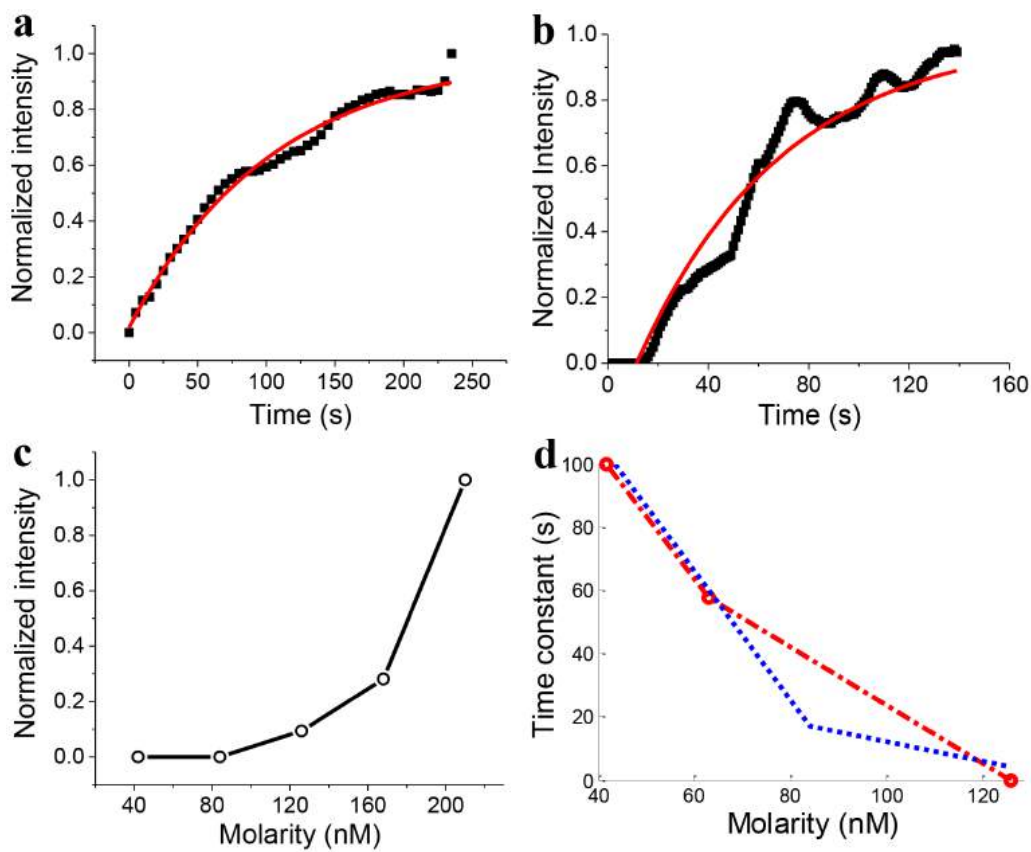

**Figure 5 | Dynamics of dye diffusion.** Increase in normalized fluorescence intensity following optoporation in the case of (a) HEK cells (84 nM) and (b) rat cortical neuron (120 nM). (c) Normalized fluorescence intensity vs RP concentration (molarity) in the case of HEK cells. Cells exhibit negative reactions at or above 168 nM concentration and are observed to die near 200 nM concentrations. (d) Theoretical (red) and experimental (blue) variation of diffusion time constant as a function of molarity.

HEK 293 cells were routinely cultured in Dulbecco's modified eagle's medium (DMEM, Sigma Aldrich), supplemented with 10% fetal bovine serum (FBS, Sigma Aldrich) and 1% Penicillin/streptomycin antibiotics. For laser-assisted insertion of rhodamine phalloidin (RP) (Cytoskeleton, Inc), cells were trypsinized and plated on poly-D-lysine coated glass-bottom 35 mm Petri dishes (MatTek Corporation).

**Dye selection and incubation.** Propidium iodide dye is widely used for demonstrating optical injection (optoporation), as it is an impermeable molecule which stains the nucleus[32–33]. Here, we have selected RP, a dye marker which is routinely used in visualizing filamentous actin (in fixed cells), and is impermeable as well as toxic at high concentrations. Our challenges were to determine an optimal concentration of RP which would allow for filamentous actin imaging, as well as ensure the cell's survival following optoporation. Determination and implementation of these parameters will allow for the dynamical study of filamentous actin polymerization and cytoskeletal reorganization in viable cells. In order to inject the appropriate amount of RP, the cell medium (DMEM) was gently removed and replaced with RP dye DMEM solution and allowed to incubate at 37°C for 15 minutes prior to optoporation. Calcein is commonly used in cell viability assays due to its retention in living cells. In order to determine the viability of cells following optoporation, Calcian-AM (4 µM) was introduced to the cell medium and allowed to incubate for 10 minutes prior to imaging.

**Optical Setup.** A schematic diagram of the experimental setup is shown in Fig. 1(a). A tunable (690–1040 nm) Ti: Sapphire laser (Newport Spectra-Physics, Inc.) beam (rep rate: 80 MHz, pulse width: ~100 fs) was directed toward the sample by a dichroic mirror (DM1) through an inverted optical microscope (Nikon Ti: U eclipse). A 100× (NA = 1.4) objective was used to focus the laser beam to a diffraction limited spot at the top surface of the cell as shown in Fig. 1(a). Special steps were taken to ensure that the laser focal plane matches with the imaging focal plane. A second dichroic mirror (DM2) was used to reflect the fluorescence excitation light from the mercury lamp along the same path as the fs laser beam. In the same filter cube as DM2, the excitation (Ex) and emission (Em) filters were used to transmit and collect the appropriate bands of visible light to and from the sample, as well as block any remaining backscattered laser light. All images were acquired by cooled EMCCD (Cascade 1 K, Photometrics) and processed with ImageJ (NIH) software. The number of fs laser pulses irradiating each sample was controlled by an external mechanical shutter (S, Uniblitz Inc). The sample-site laser beam power was controlled by the fs laser software (Mai Tai), with fine adjustments made by altering the orientation of the polarizer (P). The sample-site beam power (after the objective) was calculated by multiplying the transmission factor of the microscope objective[34] with the power measured at the back aperture of the objective. Fluorescence and bright field images were taken before and after optoporation experiments. Time lapse fluorescence images were taken immediately following optoporation in order to monitor the rise in fluorescence intensity with time.

1. Stockert, J. C., Juarranz, A., Villanueva, A. & Canete, M. Photodynamic damage to HeLa cell microtubules induced by thiazine dyes. *Cancer Chemoth. Pharm.* **39**, 167–169 (1996).
2. Jackson, S. L. & Heath, I. B. The Dynamic Behavior of Cytoplasmic F-Actin in Growing Hyphae. *Protoplasma* **173**, 23–34 (1993).
3. Sanders, M. C. & Wang, Y. L. Exogenous Nucleation Sites Fail to Induce Detectable Polymerization of Actin in Living Cells. *J. Cell Biol.* **110**, 359–365 (1990).
4. Hapala, I. Breaking the barrier: methods for reversible permeabilization of cellular membranes. *Crit. Rev. Biotechnol.* **17**, 105–122 (1997).
5. Gresch, O. et al. New non-viral method for gene transfer into primary cells. *Methods* **33**, 151–163 (2004).
6. Heiser, W. C. *Gene delivery to mammalian cells,* (Humana Press, Totowa, N.J., 2004).
7. Thomas, C. E., Ehrhardt, A. & Kay, M. A. Progress and problems with the use of viral vectors for gene therapy. *Nat. Rev. Genet.* **4**, 346–358 (2003).
8. Berns, M. W. et al. Laser microsurgery in cell and developmental biology. *Science* **213**, 505–513 (1981).
9. Tan, Y., Sun, D. & Huang, W. Mechanical modeling of red blood cells during optical stretching. *J. Biomed. Eng.* **132**, 044504 (2010).
10. Mohanty, S. Optically-actuated translational and rotational motion at the microscale for microfluidic manipulation and characterization. *Lab Chip* **12**, 3624–3636 (2012).

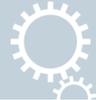

11. Gu, L., Koymen, A. R. & Mohanty, S. K. Crystalline magnetic carbon nanoparticle assisted photothermal delivery into cells using CW near-infrared laser beam. *Sci. Rep.* **4**, 5106 (2014).
12. Soughayer, J. S. *et al.* Characterization of cellular optoporation with distance. *Anal. Chem.* **72**, 1342–1347 (2000).
13. Rendall, H. A. *et al.* High-throughput optical injection of mammalian cells using a Bessel light beam. *Lab Chip* **12**, 4816–4820 (2012).
14. Marchington, R. F., Arita, Y., Tsampoula, X., Gunn-Moore, F. J. & Dholakia, K. Optical injection of mammalian cells using a microfluidic platform. *Biomed. Opt. Express* **1**, 527–536 (2010).
15. Kodama, T., Doukas, A. G. & Hamblin, M. R. Shock wave-mediated molecular delivery into cells. *BBA Mol. Cell Res.* **1542**, 186–194 (2002).
16. Kodama, T., Hamblin, M. R. & Doukas, A. G. Cytoplasmic molecular delivery with shock waves. *FASEB J.* **14**, A1473–A1473 (2000).
17. Vogel, A., Linz, N., Freidank, S. & Paltauf, G. Femtosecond-laser-induced nanocavitation in water: implications for optical breakdown threshold and cell surgery. *Phys. Rev. lett.* **100**, 038102 (2008).
18. Schaffer, C., Nishimura, N., Glezer, E., Kim, A. & Mazur, E. Dynamics of femtosecond laser-induced breakdown in water from femtoseconds to microseconds. *Opt. Express* **10**, 196–203 (2002).
19. Schneckenburger, H., Hendinger, A., Sailer, R., Strauss, W. S. L. & Schmitt, M. Laser-assisted optoporation of single cells. *J. Biomed. Opt.* **7**, 410–416 (2002).
20. Palumbo, G. *et al.* Targeted gene transfer in eucaryotic cells by dye-assisted laser optoporation. *J. Photochem. Photobiol. B* **36**, 41–46 (1996).
21. Mohanty, S. K., Sharma, M. & Gupta, P. K. Laser-assisted microinjection into targeted animal cells. *Biotechnol. Lett.* **25**, 895–899 (2003).
22. Cooper, J. A. Effects of cytochalasin and phalloidin on actin. *J. Cell. Biol.* **105**, 1473–1478 (1987).
23. Gu, L. & Mohanty, S. K. Targeted microinjection into cells and retina using optoporation. *J. Biomed. Opt.* **16**, 128003 (2011).
24. Luo, L., Callaway, E. M. & Svoboda, K. Genetic dissection of neural circuits. *Neuron* **57**, 634–660 (2008).
25. Waggoner, A. Fluorescent labels for proteomics and genomics. *Curr. Opin. Chem. Biol.* **10**, 62–66 (2006).
26. Vogel, A., Noack, J., Huttman, G. & Paltauf, G. Mechanisms of femtosecond laser nanosurgery of cells and tissues. *Appl. Phys. B-Lasers O* **81**, 1015–1047 (2005).
27. Tien, A. C., Backus, S., Kapteyn, H., Murnane, M. & Mourou, G. Short-pulse laser damage in transparent materials as a function of pulse duration. *Phys. Rev. Lett.* **82**, 3883–3886 (1999).
28. Gordienko, V. M., Mikheev, P. M. & Syrtsov, V. S. Nonmonotonic behavior of the absorption of the tightly focused femtosecond radiation of a Cr: Forsterite laser in a dielectric due to an increase in the number of photons involved in the process. *JETP Lett.* **82**, 228–231 (2005).
29. Born, M. & Wolf, E. *Principles of optics: electromagnetic theory of propagation, interference and diffraction of light,* (Cambridge University Press, Cambridge; New York, 1999).
30. Grill, S. & Stelzer, E. H. K. Method to calculate lateral and axial gain factors of optical setups with a large solid angle. *J Opt. Soc. Am. A* **16**, 2658–2665 (1999).
31. Parker, W. C. *et al.* High-resolution intracellular viscosity measurement using time-dependent fluorescence anisotropy. *Opt. Express* **18**, 16607–16617 (2010).
32. Baumgart, J. *et al.* Quantified femtosecond laser based opto-perforation of living GFSHR-17 and MTH53 a cells. *Opt. Express* **16**, 3021–3031 (2008).
33. Davis, A. A., Farrar, M. J., Nishimura, N., Jin, M. M. & Schaffer, C. B. Optoporation and genetic manipulation of cells using femtosecond laser pulses. *Biophys. J.* **105**, 862–871 (2013).
34. Kong, X. *et al.* Comparative analysis of different laser systems to study cellular responses to DNA damage in mammalian cells. *Nucleic Acids Res.* **37**, e68 (2009).

### Acknowledgments

The authors would like to thank Prof. Young-tae Kim for preparation of cortical neuron samples. SM would like to thank the supports from the National Institute of Health (NS084311), National Science Foundation (1148541) and Office of President and Provost, The University of Texas at Arlington.

### Author contributions

S.M. conceived and supervised the project. K.D. performed experiments, analyzed data and wrote the paper. B.B. assisted K.D. doing the experiments and writing paper. All authors reviewed the manuscript.

### Additional information

**Supplementary information** accompanies this paper at http://www.nature.com/scientificreports

**Competing financial interests**: The authors declare no competing financial interests.

**How to cite this article**: Dhakal, K., Black, B. & Mohanty, S. Introduction of impermeable actin-staining molecules to mammalian cells by optoporation. *Sci. Rep.* **4**, 6553; DOI:10.1038/srep06553 (2014).